\documentclass[11pt]{article}
\usepackage{amsfonts, amsmath, amssymb, amsthm}
\usepackage{graphicx}
\usepackage[margin=1in]{geometry}
\usepackage{natbib}
\usepackage{setspace}
\usepackage{enumerate}
\usepackage{enumitem}
\usepackage{mathtools}
\usepackage[toc,title,titletoc,header]{appendix}
\usepackage[colorlinks,citecolor=blue,urlcolor=blue, linkcolor=blue]{hyperref}
\usepackage{etoolbox} 
\usepackage{bbm}
\usepackage{xcolor}

\def\qed{\rule{2mm}{2mm}}
\def\indep{\perp \!\!\! \perp}

\let\footnote=\endnote

\usepackage[pagewise,mathlines]{lineno}
\mathchardef\dash="2D

 \definecolor{purple}{rgb}{0.7, 0.16, 1}

\newcommand{\tr}{r}
\newtheorem{theorem}{Theorem}[section]
\newtheorem{lemma}{Lemma}[section]

\theoremstyle{definition}

\newtheorem{remark}{Remark}[section]
\newtheorem{assumption}{Assumption}[section]
\newtheorem{algorithm}{Algorithm}[section]

\newcommand{\tmax}{t_{\max}}

\newcommand{\Real}{\mathbf{R}}

\newcommand{\Ynt}{Y^{(n)}}

\newcommand{\Xn}{X^{(n)}}
\newcommand{\Tn}{T^{(n)}}

\newcommand{\hin}{\hat Y_{I_1, t}}

\AtEndEnvironment{remark}{~\qed}
\AtEndEnvironment{example}{~\qed}

\newcommand{\Ind}{\mathbbm{1}}

\begin{document}

\author{
Azeem M.\ Shaikh\\
Department of Economics\\
University of Chicago \\
\url{amshaikh@uchicago.edu}
\and
Panos Toulis \\
Booth School of Business \\
University of Chicago \\
\url{ptoulis@chicagobooth.edu}
}

\bigskip

\title{Randomization Tests in Observational Studies with\\ Staggered Adoption of Treatment\thanks{The research of the first author is supported by NSF Grant SES-1530661.}}

\maketitle

\vspace{-0.9cm}

\begin{spacing}{1.1}
\begin{abstract}
This paper considers the problem of inference in observational studies with time-varying adoption of treatment. In addition to an unconfoundedness assumption that the potential outcomes are independent of the times at which units adopt treatment conditional on the units' observed characteristics, our analysis assumes that the time at which each unit adopts treatment follows a Cox proportional hazards model. This assumption permits the time at which each unit adopts treatment to depend on the observed characteristics of the unit, but imposes the restriction that the probability of multiple units adopting treatment at the same time is zero. In this context, we study randomization tests of a null hypothesis that specifies that there is no treatment effect for all units and all time periods in a distributional sense.  We first show that an infeasible test that treats the parameters of the Cox model as known has rejection probability under the null hypothesis no greater than the nominal level in finite samples.  Since these parameters are unknown in practice, this result motivates a feasible test that replaces these parameters with consistent estimators.  While the resulting test does not need to have the same finite-sample validity as the infeasible test, we show that it has limiting rejection probability under the null hypothesis no greater than the nominal level. In a simulation study, we examine the practical relevance of our theoretical results, including robustness to misspecification of the model for the time at which each unit adopts treatment.  Finally, we provide an empirical application of our methodology using the synthetic control-based test statistic and tobacco legislation data found in \citet{abadie2010synthetic}.
\end{abstract}
\end{spacing}



\thispagestyle{empty} 
\newpage
\setcounter{page}{1}

\section{Introduction} \label{sec:intro}

This paper considers the problem of inference in observational studies in which units adopt treatment at varying times and remain treated once adopting treatment.  The widespread availability of data with this type of structure has led to the development of several methods for its analysis, including difference-in-differences \citep{snow1855mode,card1993minimum} and, more recently, synthetic controls \citep{abadie2003economic,abadie2010synthetic,abadie2019using}.  For a modern overview, see Section 5 of \cite{abadie2018econometric}; further references are provided below.  In contrast to the literature on difference-in-differences, which has focused largely on the estimation of average effects of the treatment on the outcome of interest, we study randomization tests of a null hypothesis that specifies the treatment has no effect on the outcome of interest for all units and all time periods in a distributional sense.  This null hypothesis is related to, but less restrictive than the more conventional ``sharp'' null hypothesis that the treatment has no effect on the outcome of interest for all units and all time periods.  This latter null hypothesis has appeared previously in the literature on synthetic controls \citep{abadie2010synthetic, firpo2018synthetic}, but, in contrast to this literature, which has focused primarily on settings in which there is only a single unit that adopts treatment, our testing procedure exploits the availability of multiple units that adopt treatment at different times in a novel way that we describe further below.  

In addition to an unconfoundedness assumption that the potential outcomes are independent of the times at which units adopt treatment conditional on observed characteristics, the main restriction underlying our analysis is a survival model for the time at which each unit adopts treatment. We require, in particular, that the time at which each unit adopts treatment follows a Cox proportional hazards model. This assumption permits the time at which each unit adopts treatment to depend on the observed characteristics of the unit, but restricts the probability of multiple units adopting treatment at the same time to be zero. For this reason, as discussed further in Remark \ref{remark:ties} below, we view our methodology as being best suited for settings in which time is measured with sufficient granularity to make this probability small.  A key consequence of the Cox proportional hazards assumption for our purposes is that it implies a parametric restriction on the distribution of the identity of the unit that first adopted treatment conditional on observed characteristics and the time of such first treatment adoption.  As described further in Remark \ref{remark:designvsmodel}, we view our analysis as ``design-based'' in the sense that it exploits heavily this more subtle implication of our assumption on the distribution of the times at which each unit adopts treatment.  Even though the Cox proportional hazards model is semiparametric in nature, we emphasize that it is restrictive.  For that reason, in the spirit of the ``double robustness'' literature, we further provide some results pertaining to the robustness of our procedure to violations of this assumption in Section \ref{sec:robustness}.

In order to motivate our proposed testing procedure, our first result shows that an infeasible test that treats the parameters of this conditional distribution of the identity of the unit that first adopted treatment --- more succinctly, the ``first adopter" --- as known has rejection probability under the null hypothesis no greater than the nominal level in finite samples.  We then establish that the feasible test that replaces these parameters with consistent estimators has limiting rejection probability under the null hypothesis no greater than the nominal level.  We emphasize that consistent estimation of these parameters relies upon the availability of multiple units that adopt treatment at different times.  As mentioned previously, this feature distinguishes our analysis from some analyses found in the literature on synthetic controls in which there is only a single treated unit.  We also note that our method for establishing this limiting result is novel in that it relies upon results in \cite{romano2012uniform} to argue that the difference in the rejection probabilities of the feasible and infeasible tests tends to zero as the number of units becomes large.  As explained further in Remark \ref{remark:np}, this technique permits one to establish a nonparametric counterpart to our procedure under appropriate conditions, but we expect that such a procedure may perform poorly in finite samples for even moderately many observed unit characteristics.

Our analysis is most closely related to randomization tests in the literature on synthetic controls.  If the conditional distribution of the identity of the first adopter in our analysis is uniform, then our testing procedure reduces to the randomization test proposed by \cite{abadie2010synthetic}.  In the same spirit as the literature on sensitivity analysis in observational studies \citep{rosenbaum1987sensitivity,rosenbaum2002observational}, \cite{firpo2018synthetic} parameterize this conditional distribution in order to explore the sensitivity of the testing procedure to deviations from the assumption that it is uniform.  The parameterization used by \cite{firpo2018synthetic} resembles our expression for the conditional distribution of the identity of the first adopter.  In this way, a by-product of our analysis is an alternative viewpoint on their specific parameterization.  \cite{chernozhukov2017exact} have also recently proposed randomization tests for synthetic controls, but their analysis exploits exchangeability assumptions across the temporal dimension, whereas we, as explained previously, exploit a cross-sectional restriction of our survival model on the conditional distribution of the identity of the first adopter.   Other recent proposals for inference in the context of synthetic controls include \cite{cattaneo2019synth}, who propose methods based on concentration inequalities from the high-dimensional statistics literature, and \cite{li2019statistical}, who propose methods based on subsampling that are applicable whenever both the number of pre-treatment and number of post-treatment time periods are large.

The remainder of the paper is organized as follows.  In Section \ref{sec:setup}, we describe our setup and notation.  In particular, there we describe the Cox proportional hazards model that we use to model the time at which units adopt treatment and the resulting parametric restriction on the distribution of the identity of the first adopter unit conditional on observed characteristics, and the time of first treatment adoption.  Section \ref{sec:main} contains our main results, beginning with the finite-sample result for the infeasible test before presenting the large-sample result for the feasible test.  In Section \ref{sec:impl}, we summarize the steps required to implement our proposed test and discuss some related issues, including the robustness of our testing procedure to violations of the Cox proportional hazards model. We explore the practical relevance of our theoretical results via a simulation study in Section \ref{sec:sims}.  Finally, in Section \ref{sec:empirical}, we provide an empirical application of our methodology using the synthetic control-based test statistic and tobacco legislation data found in \cite{abadie2010synthetic}.  Section \ref{sec:conclusion} concludes briefly.  Proofs of all results can be found in the Appendix.

\section{Setup and Notation} \label{sec:setup}

We index units by $i \in \mathbb N = \{1, \ldots, n\}$ and time by $t \in [0,\infty)$.  Denote by $X_{i,t} \in \Real^d$ the characteristics of unit $i$ at time $t$ and by $T_i \in [0,\infty)$ the time at which unit $i$ adopts treatment. Each unit may adopt treatment only once, and remains treated thereafter. Define $Y_{i,t}(\tr)$ to be the potential outcome of unit $i$ at time $t$ if the time at which treatment had been adopted was equal to $\tr \in [0, \infty)$.  This framework permits potential outcomes at time $t$ to depend not only on whether treatment was adopted earlier or later than $t$, i.e., $\Ind\{\tr \leq t\}$, but also on the time at which treatment was adopted or the time since treatment was adopted.  The outcome of unit $i$ at time $t$ is therefore given by 
\begin{equation} \label{eq:obsy}
Y_{i,t} = Y_{i,t}(T_i)~.
\end{equation}
We denote by $\mathbb T = \{1, \ldots, \tmax\}$ the subset of times at which units' characteristics and outcomes are observed. In what follows, we will sometimes require $X_{i,t}$ to be defined for values of $t \not \in \mathbb T$.  For such values, it is understood that $X_{i,t}$ equals $X_{i,[t]}$, where $[t]$ maps $t \in [0,\infty)$ to the ``closest'' value in $\mathbb T$.  For example, 
\begin{equation*}
[t] =
\begin{cases}
\lceil t \rceil & \text { if } t \leq \tmax \\
\tmax & \text { if } t > \tmax
\end{cases}
~.
\end{equation*}
We emphasize, however, that our analysis will not make use of values of $Y_{i,t}$ for $t \not \in \mathbb T$.

We adopt the following shorthand notation:
\begin{eqnarray*}
\Ynt(\tr^{(n)}) &=& (Y_{i,t}(\tr_i) : i \in \mathbb N, t \in \mathbb T ) \\
\Ynt &=& (Y_{i,t} : i \in \mathbb N, t \in \mathbb T ) \\
X_i &=& (X_{i,t} : t \in \mathbb T) \\
\Xn &=& (X_{i,t} : i \in \mathbb N, t \in \mathbb T) \\
\Tn &=& (T_i : i \in \mathbb N)~,
\end{eqnarray*}
where $\tr^{(n)} =  (\tr_i : i \in \mathbb N) \in [0,\infty)^n$. Note that $\Ynt$ and $\Xn$ are observed, but adoption times may be censored. Indeed, $T_i$ is not observed whenever $T_i > \tmax$.  In what follows, we also make use of $$T_{(1)} = \min_{i \in \mathbb N} T_i~,$$ the time at which a unit first adopts treatment, as well as $$I_1 = \arg \min_{i \in \mathbb N} T_i~,$$ the  (random) index corresponding to the first adopter.  Note that $I_1$ may be defined in this way provided that ties occur with probability zero, which will be ensured by Assumption \ref{ass:prop} below.

Using the notation introduced above, we may formally state the null hypothesis of interest as
\begin{equation} \label{eq:null}
H_0 : \Ynt(\tr^{(n)}) | X^{(n)} \stackrel{d}{=} \Ynt(\tr^{\prime (n)}) | X^{(n)} \text{ for all } \tr^{(n)}, \tr^{\prime (n)} \in [0,\infty)^n~.
\end{equation}
This null hypothesis specifies a sense in which the treatment has no effect on potential outcomes for all units and all time periods that is closely related to but less restrictive than the ``sharp'' null hypothesis that specifies $\Ynt(\tr^{(n)}) = \Ynt(\tr^{\prime (n)})$ for all $\tr^{(n)}, \tr^{\prime (n)} \in [0,\infty)^n$ with probability one.   As mentioned previously, this latter type of null hypothesis has already appeared in the literature on synthetic controls \citep{abadie2010synthetic,firpo2018synthetic}.

Our analysis will require the following assumptions.
\begin{assumption} \label{ass:unconf} ({\it Unconfoundedness})
$(\Ynt(\tr^{(n)}) : \tr^{(n)} \in [0,\infty)^n) \indep \Tn \big | \Xn$.
\end{assumption}

\noindent A central object in our analysis will be the conditional distribution
\begin{equation} \label{eq:conddistr}
I_1 \big | T_{(1)}, \Xn.
\end{equation}
This distribution  will be governed by the following additional assumption:
\begin{assumption} \label{ass:prop} ({\it Proportional Hazards Model}) The distribution of $(T^{(n)}, X^{(n)})$ is such th{}at 
\begin{itemize}[topsep=0pt,itemsep=1pt]
\item[(a)] $T_1, \ldots, T_n$ are independent conditional on $X^{(n)}$.
\item[(b)] $T_i|X^{(n)} \stackrel{d}{=} T_i | X_i$.
\item[(c)] $T_i|X_i$ has density w.r.t.\ Lebesgue measure such that, for some $\beta \in \mathbb B \subseteq \Real^d$,
\begin{equation} \label{eq:hazardprop}
\lim_{\delta \downarrow 0} \frac{1}{\delta} P\left \{t \leq T_i \leq t + \delta \big | T_i \geq t, X_i \right \} = \lambda(t)\exp(X_{i,t}'\beta).
\end{equation}
\item[(d)] $\text{supp}(X_i)$ is bounded.
\end{itemize}
\end{assumption}
\noindent Our final assumption simply stipulates that we have access to a consistent estimator, $\hat \beta_n$, of $\beta$. 
\begin{assumption} \label{ass:cons} ({\it Consistent Estimator}) There exists a consistent estimator $\hat \beta_n$ of $\beta$, i.e., $$\hat \beta_n \stackrel{P}{\rightarrow} \beta~.$$
\end{assumption}
\noindent In a setting governed by Assumption \ref{ass:prop}, it is natural to define $\hat \beta_n$ to be 
\begin{equation} \label{eq:hatbeta}
\hat \beta_n = \arg \max_{\beta \in \mathbb B} \prod_{1 \leq i \leq n} \left ( \frac{\exp(X_{i,T_i}'\beta)}{\sum_{j \in \mathcal R_i} \exp(X_{j,T_j}'\beta)} \right )^{1 - \delta_i}~,
\end{equation}
where $\delta_i = \Ind\{T_i > \tmax\}$ is an indicator for censorship of $T_i$ and $\mathcal R_i = \{j \in \mathbb N : T_j \geq T_i\}$ is, in the language of the survival analysis literature, the ``risk set'' at time $T_i$, i.e., the set of units that have not yet adopted treatment at time $T_i$. We note that the maximand on the right-hand side of \eqref{eq:hatbeta} is the celebrated partial likelihood of \cite{cox1975partial}.  Sufficient conditions for Assumption \ref{ass:cons} when $\hat \beta_n$ is defined by \eqref{eq:hatbeta} can be found, for example, in \cite{andersen1982cox}.   

An important special case of our framework is one in which the potential outcomes measured at some time $t$ are only permitted to depend on whether treatment was already adopted at $t$, i.e., 
\begin{equation} \label{eq:restrictive}
Y_{i,t}(\tr) = \tilde Y_{i,t}(0) + \Ind\{\tr \leq t\}(\tilde Y_{i,t}(1) - \tilde Y_{i,t}(0))~,
\end{equation}
where $\tilde Y_{i,t}(1)$ and $\tilde Y_{i,t}(0)$ the potential outcomes of unit $i$ at time $t$ under treatment and control, respectively.  In this formulation, the null hypothesis in \eqref{eq:null} becomes $$\tilde H_0 : \tilde Y^{(n)}(d^{(n)}) | X^{(n)} \stackrel{d}{=} \tilde Y^{(n)}(d^{\prime (n)}) | X^{(n)}~,$$ where $$\tilde Y^{(n)}(d^{(n)}) = (\tilde Y_{i,t}(d_i) : i \in \mathbb N, t \in \mathbb T)$$ and $d^{(n)} = (d_i : i \in \mathbb N) \in \{0,1\}^n$.  If it is assumed further that treatment effects are constant, i.e., 
\begin{equation} \label{eq:const}
\tilde Y_{i,t}(1) - \tilde Y_{i,t}(0) = \tau \text{ for some constant } \tau~,
\end{equation}
then $\tilde H_0$ simplifies further to testing $\tau = 0$.  If both \eqref{eq:restrictive} and \eqref{eq:const} are assumed, then it is straightforward to modify our testing procedure below to test the null hypothesis that $\tau = \tau_0$ for some pre-specified $\tau_0\in\Real$. Indeed, all that is required is simply to replace $Y_{i,t}$ with $Y_{i,t} - \tau_0 \Ind\{T_i \leq t\}$.  In this way, a confidence region for $\tau$ may be constructed by inverting such hypothesis tests.

\begin{remark} \label{remark:ties}
({\it Tied adoption times}). As mentioned previously, an implication of Assumption \ref{ass:prop} for our purposes is that the probability of multiple units adopting treatment at the same time is zero.  Since in most empirical applications the time of treatment adoption is only measured discretely, we view our methodology as being suited to settings in which treatment adoption times are measured with sufficient granularity so that this probability is small.  In our empirical application in Section \ref{sec:empirical}, for example, $T_i$ denotes the time at which a state adopts tobacco legislation and our data permit measurement up to the month or even the day of adoption.
\end{remark}

\section{Main Results} \label{sec:main}

In order to motivate our proposed testing procedure, it is useful first to describe an infeasible test of \eqref{eq:null} that assumes $\beta$ is known and is level $\alpha$ in finite samples.  To this end, denote by 
\begin{equation}\label{eq:tstat}
S_n^{\rm obs} = S_n\left (I_1, \Ynt, T_{(1)}, \Xn \right )
\end{equation}
a test statistic such that large values of $S_n^{\rm obs}$ provide evidence against \eqref{eq:null}.  While \eqref{eq:tstat} imposes some restrictions on the form of the test statistic, it accommodates many test statistics used in the literature, including choices in \cite{abadie2010synthetic} and \cite{firpo2018synthetic}.  See Section \ref{sec:choice} below for further discussion.  In order to describe a suitable critical value with which to compare $S_n^{\rm obs}$, we begin with a straightforward implication of the null hypothesis \eqref{eq:null} and Assumption \ref{ass:unconf}.  

\begin{lemma} \label{lemma:indep}
Under Assumption \ref{ass:unconf}, 
\begin{equation} \label{eq:indep}
\Ynt \indep I_1 \big | T_{(1)}, X^{(n)},
\end{equation}
whenever the null hypothesis \eqref{eq:null} holds.
\end{lemma}

\noindent We also note the following characterization of \eqref{eq:conddistr}, the conditional distribution of the identity of the first adopter, $I_1$, under Assumption \ref{ass:prop}:

\begin{lemma} \label{lemma:prop}
Under Assumption \ref{ass:prop}, $P\{I_1 = i | T_{(1)}, \Xn\} = \omega_{i, n}(\beta)$, where
\begin{equation} \label{eq:pin}
\omega_{i, n}(\beta) = \frac{\exp(X_{i,T_{(1)}}'\beta)}{\sum_{1 \leq k \leq n} \exp(X_{k,T_{(1)}}'\beta)}~.
\end{equation}
\end{lemma}

\noindent It is possible to use these two lemmas to calculate the distribution of $S_n^{\rm obs}$ conditional on $\Ynt, T_{(1)}, \Xn$ exactly whenever the null hypothesis \eqref{eq:null} holds.  Indeed, whenever \eqref{eq:null} holds, 
\begin{equation} \label{eq:cdf}
P\left \{S_n^{\rm obs} \leq s \big | \Ynt, T_{(1)}, \Xn \right \} = 
\sum_{1 \leq i \leq n} \omega_{i,n}(\beta) \Ind\left \{S_n\left (i, \Ynt, T_{(1)} , \Xn \right ) \leq s \right \}~,
\end{equation}
where $\omega_{i,n}(\beta)$ is given by \eqref{eq:pin}.  A suitable critical value with which to compare $S_n^{\rm obs}$ is therefore given by 
\begin{equation} \label{eq:crit}
\hat c_n(1 - \alpha, \beta) = \inf \left \{ s \in \mathbf R : \sum_{1 \leq i \leq n} \omega_{i,n}(\beta) 
\Ind\left \{S_n\left (i, \Ynt, T_{(1)}, \Xn \right ) \leq s \right \} \geq 1 - \alpha \right \}~. 
\end{equation}
By construction, the test of \eqref{eq:null} that rejects the null hypothesis if and only if $S_n^{\rm obs}$ exceeds $\hat c_n(1 - \alpha, \beta)$ is level $\alpha$ in finite samples. 
This test is, of course, infeasible because $\beta$ is unknown. Our feasible test is given by replacing $\beta$ with a consistent estimator, $\hat \beta_n$.  The following theorem shows that, under our assumptions, the resulting test, i.e., 
\begin{equation} \label{eq:test}
\phi_n = \Ind\left \{S_n^{\rm obs} > \hat c_n(1 - \alpha, \hat \beta_n) \right \}~,
\end{equation} 
has limiting rejection probability under the null hypothesis no greater than the nominal level.  Note that in order to compute $\omega_{i,n}(\hat \beta_n)$ and thus the test defined in \eqref{eq:test} we require $T_{(1)} < \tmax$, which occurs with probability approaching one under our assumptions.

\begin{theorem} \label{thm:main}
If Assumptions \ref{ass:unconf}--\ref{ass:cons} hold, then the test $\phi_n$ defined in \eqref{eq:test} satisfies $$\limsup_{n \rightarrow \infty} E[\phi_n] \leq \alpha$$ whenever the null hypothesis \eqref{eq:null} holds.
\end{theorem}

The proof of Theorem \ref{thm:main} involves showing that the difference in the rejection probabilities of the infeasible and feasible tests tends to zero as $n$ tends to infinity.  The desired result then follows immediately since it is known that the infeasible test has rejection probability under the null hypothesis no greater than the nominal level in finite samples.  A key step in the argument is to use Lemma A.1 in \cite{romano2012uniform} to show that the difference in these rejection probabilities may be linked to the sum of the differences in $\omega_{i,n}(\beta)$ and $\omega_{i,n}(\hat \beta_n)$.  Even though this bound involves a growing number of terms, we show that it is possible to control it using a combination of the boundedness of the support of $X_{i,t}$ and the consistency of $\hat \beta_n$.

We note that a $p$-value corresponding to the test \eqref{eq:test} may be defined as 
\begin{equation} \label{eq:truepval}
\inf \{ \alpha \in (0,1) : S_n^{\rm obs} > \hat c_n(1 -\alpha, \hat \beta_n) \}~.
\end{equation}
In order to facilitate computation, it is useful to define 
\begin{equation} \label{eq:pval}
\hat p_n(\omega) = \sum_{1 \leq i \leq n} \omega_i \Ind\{S_n(i,\Ynt, T_{(1)}, \Xn )\geq S_n^{\rm obs}\}~,
\end{equation}
where $S_n^{\rm obs}$ is defined as in \eqref{eq:tstat} and $\omega$ is an element of the $n$-dimensional simplex. 
In this notation, the $p$-value of our test defined in \eqref{eq:truepval} is simply given by \eqref{eq:pval} with $\omega = (\omega_{i,n}(\hat \beta_n) : i \in \mathbb N)$.  The $p$-values corresponding to other tests described above may be computed in a similar way.  For example, the $p$-value of the infeasible test of \eqref{eq:null} that rejects the null hypothesis if and only if $S_n^{\rm obs}$ exceeds $\hat c_n(1  - \alpha,\beta)$ rather than $\hat c_n(1  - \alpha,\hat \beta_n)$ is given by \eqref{eq:pval} with $\omega = (\omega_{i,n}(\beta) : i \in \mathbb N)$.

\begin{remark} \label{remark:designvsmodel}
({\it Model-based vs.\ design-based inference}). Following \cite{athey2018design}, it is common to describe some methods for inference in settings where the distribution of treatment is known, such as randomized controlled trials, as ``design-based'' rather than ``model-based.''  This terminology is intended to distinguish inference methods that exploit (only) information about the ``design,'' i.e., the distribution of treatment, from methods that (additionally) exploit information about the distribution of outcomes.  The test defined in \eqref{eq:test} conditions on $Y^{(n)}$, $T_{(1)}$ and $X^{(n)}$ and exploits only variation in the identity of the first adopter, which is governed by the distribution of treatment specified in Assumption \ref{ass:prop}.  In this sense, we view our test as being ``design-based'' rather than ``model-based.''
\end{remark}

\begin{remark} 
({\it Randomized vs.\ non-randomized tests}). The test defined in \eqref{eq:test} is non-randomized in that rejects if and only if $S_n^{\rm obs} > \hat c_n(1 - \alpha, \hat \beta_n)$.  It is possible to define the test so that it is randomized in that it additionally rejects with probability $q$ when $S_n^{\rm obs} = \hat c_n(1 - \alpha, \hat \beta_n)$, where 
$$q = (1 - \alpha) - \sum_{1 \leq i \leq n} \omega_{i,n}(\hat \beta_n) \Ind\{S_n(i,\Ynt, T_{(1)}, \Xn ) \leq \hat c_n(1 - \alpha,\hat \beta_n)\}~.$$ 
Similar modifications are often used in the context of related tests to achieve exactness and may have an especially noticeable effect when $n$ is small.  See \citet[Section 15.2.1]{lehmann2006testing} for further discussion.  Here, despite its apparent similarity, we emphasize that the uses of the adjectives `randomized' and `non-randomized' are distinct from the use of the term `randomization' in the description of randomization tests.
\end{remark}

\begin{remark} \label{rem:multipletimes}
({\it Multiple adoption times}). It is natural to consider tests that condition not just on $T_{(1)}$, but $T_{(1)}$ and $T_{(2)}$, where $T_{(2)}$ is the second-order statistic of $T_1, \ldots, T_n$.  By arguing as in the proof of Lemma \ref{lemma:prop}, it is possible to see that $P\{I_1 = i, I_2 = j | T_{(1)}, T_{(2)}, X^{(n)}\}$ depends not only on $(X_{i,t}: i \in \mathbb N, t \in \{T_{(1)},T_{(2)}\})$, but also on the integral of $\lambda(t)\exp(X_{i,t}'\beta)$ over $T_{(1)} \leq t \leq T_{(2)}$.  For this reason, we do not pursue such tests further in this paper.
\end{remark}

\begin{remark} \label{remark:np}
({\it Nonparametric tests}). While we believe it would not be prudent to do so, it is possible to adapt our test for use with a nonparametric estimate of the left-hand side of \eqref{eq:hazardprop}.  Indeed, if we replace the right-hand side of \eqref{eq:hazardprop} in Assumption \ref{ass:prop} with $\lambda(t|X_{i,t})$, then, by arguing as in the proof of Lemma \ref{lemma:prop}, it is possible to deduce that $$P\{I_i = i | T_{(1)}, X^{(n)}\} = \frac{\lambda(T_{(1)}|X_{i,T_{(1)}})}{\sum_{1 \leq k \leq n}\lambda(T_{(1)}|X_{k,T_{(1)}})}~.$$  Consider the test defined in \eqref{eq:test}, but in which $\omega_{i,n}(\beta)$ defined in \eqref{eq:crit} is replaced with $$\frac{\hat \lambda_n(T_{(1)}|X_{i,T_{(1)}})}{\sum_{1 \leq k \leq n}\hat \lambda_n(T_{(1)}|X_{k,T_{(1)}})}~,$$ where $\hat \lambda_n(t|X_{i,t})$ is an estimator of $\lambda(t|X_{i,t})$.  By inspecting the proof of Theorem \ref{thm:main}, we see that this test has limiting rejection probability under the null hypothesis \eqref{eq:null} no greater than the nominal level whenever 
\begin{equation} \label{eq:npass}
\sup_{1 \leq i \leq n} \left | \frac{\lambda(T_{(1)}|X_{i,T_{(1)}})}{\frac{1}{n}\sum_{1 \leq k \leq n}\lambda(T_{(1)}|X_{k,T_{(1)}})} - \frac{\hat \lambda_n(T_{(1)}|X_{i,T_{(1)}})}{\frac{1}{n} \sum_{1 \leq k \leq n}\hat \lambda_n(T_{(1)}|X_{k,T_{(1)}})} \right | \stackrel{P}{\rightarrow} 0~.
\end{equation}
Using results from the literature on nonparametric estimation of conditional hazard rates, it is possible to provide more primitive conditions under which \eqref{eq:npass} holds; see, for example, \cite{spierdijk2008nonparametric}.  We, however, refrain from doing so because with covariates of even moderate dimensionality we believe such an approach may not behave well in finite samples. 
\end{remark}

\section{Implementation and Related Discussion} \label{sec:impl}

In this section, we summarize the steps required to implement our proposed testing procedure and provide some discussion of related issues.  To that end, we first present the following algorithmic description of the test defined in \eqref{eq:test}:
\begin{algorithm} \hspace{1cm}\label{algo1}
\begin{enumerate}
\item[] {\bf Step 1}:  Use $(T^{(n)},X^{(n)})$ to compute $\hat \beta_n$ defined in \eqref{eq:hatbeta}.  In {\tt R}, one may use the package {\tt survival}; in {\tt Stata}, the same functionality is provided by the {\tt stcox} command.
\item[] {\bf Step 2}: For $i \in \mathbb N$, compute $\omega_{i,n}(\hat \beta_n)$ using the relationship in \eqref{eq:pin} with $\beta = \hat \beta_n$.
\item[] {\bf Step 3}: For a given choice of $S_n$, compute $\hat p_n(\omega)$ defined in \eqref{eq:pval} with $\omega = (\omega_{i,n}(\hat \beta_n) : i \in \mathbb N)$.
\end{enumerate}
\end{algorithm}
\noindent We note that it may be further desirable to assess the validity of the Cox proportional hazards model using various diagnostic tests.  For a recent survey of such methods, see 
\cite{xue2017diagnostics}, who further provide references to the relevant {\tt R} packages. In the subsequent subsections we (i) discuss the choice of $S_n$; (ii) provide some results related to the robustness of our procedure to violations of the Cox proportional hazards model; and (iii) relate our proposed testing procedure to two other closely related procedures in the literature.

\subsection{Choice of \texorpdfstring{$S_n$}{}} \label{sec:choice}

While our theory applies to any choice of test statistic $S_n$ that can be written as in \eqref{eq:tstat}, some choices of test statistics may be preferable in terms of power of the resulting 
test.  \cite{abadie2010synthetic} suggest, for example, a test statistic of the form
\begin{equation}\label{eq:synth}
\frac{\sum_{t\in \mathbb T : t \ge T_{(1)}}  (Y_{I_1,t}- 
\hin)^2 }
{\sum_{t\in \mathbb T : t < T_{(1)}}  (Y_{I_1,t}- \hat 
Y_{I_1, t})^2 }~,
\end{equation}
where $\hin$ is a linear combination of $\{Y_{i,t} : i \in \mathbb N \setminus\{I_1\}\}$, where the weights are chosen so that $Y_{I_1, t} \approx \hin$ for $t\in \mathbb T$ with $t < T_{(1)}$. Under the sharp null hypothesis, $\hin$ may therefore be viewed as an estimator of $Y_{I_1, t}(0)$ for $t\in \mathbb T$ with $t \geq T_{(1)}$.  The weights in the construction of $\hin$ represent the combination of the other units that are intended to be used as a control for $I_1$, i.e., the ``synthetic control.''  Different choices for these weights have been suggested by a variety of authors in the literature.  Section 2 of \cite{cattaneo2019synth} provides a succinct summary of proposals by \cite{abadie2010synthetic}, \cite{hsiao2012panel}, \cite{doudchenko2016balancing}, \cite{chernozhukov2018practical}, \cite{ferman2019synthetic}, and \cite{arkhangelsky2019synthetic}.  For further discussion of and contributions to the synthetic control literature, see \cite{abadie2017penalized}, \cite{amjad2018robust}, \cite{athey2018matrix} and \cite{ben2019synthetic}. Of course, other choices for $S_n$ are possible, including a simple difference-in-differences test statistic or a $t$-test statistic. \citet{firpo2018synthetic} explore the power of tests stemming from different choices of test statistic via a simulation study with a state space model in the specific case of our test where \eqref{eq:conddistr}, the conditional distribution of the identify of the first adopter, is uniform. They find that the test statistic in~\eqref{eq:synth} performs well both in terms of power and with respect to their sensitivity analysis, which we elaborate on further in Section \ref{sec:firpo} below.

\newcommand{\snn}{s_n^{(n)}}
\subsection{Robustness to Misspecification} \label{sec:robustness}

While the Cox proportional hazards model is widely used and semiparametric in nature, it is nevertheless natural to be concerned about the extent to which inferences made using our proposed test are robust to misspecification of the model for $T^{(n)}|X^{(n)}$ in Assumption \ref{ass:prop}.  We therefore now describe a sense in which our test is robust to this sort of misspecification. To this end, let $s_{n,i} = S_n(i, Y^{(n)},T_{(1)}, X^{(n)})$, $\snn = (s_{n,i} : i \in \mathbb N)$, and $$P_{n,i} = P\{s_{n,i} > \hat c_n(1-\alpha, \hat\beta_n) | \Tn, \Xn\}~.$$ 

Suppose there is a sequence of indices $j_n \in \mathbb N$, such that
\begin{equation}\label{eq:delta_n}
\delta_{j_n}(\Tn, \Xn) = \max_{i\in \mathbb N} |P_{n,i} - P_{n,j_n} | = o(1)
\end{equation}
whenever the null hypothesis \eqref{eq:null} holds.  Under condition~\eqref{eq:delta_n}, it follows  that $\phi_n$ defined in \eqref{eq:test} has rejection probability no greater than the nominal level under the null hypothesis even when Assumption \ref{ass:prop} fails to hold.  To see this, first note that the definition of $\hat c_n(1 - \alpha, \hat \beta_n)$ in \eqref{eq:crit} implies that $$\sum_{1 \leq i \leq n} \omega_{i,n}(\hat \beta_n) \Ind\{s_{n,i} > \hat c_n(1 - \alpha, \hat \beta_n)\} \leq \alpha~,$$ 
and so
\begin{equation} \label{eq:exch}
\sum_{1 \leq i \leq n} \omega_{i,n}(\hat \beta_n)P_{n, i} \le \alpha~.
\end{equation}
For $j_n$ satisfying \eqref{eq:delta_n} and $\underline i_n = \arg \min_{i \in \mathbb N} P_{n,i}$, we have that $$\sum_{1 \leq i \leq n} \omega_{i,n}(\hat \beta_n)P_{n, i} \geq P_{n,\underline i_n} = 
P_{n,j_n} + P_{n,\underline i_n} - P_{n, j_n} \geq P_{n,j_n} - \delta_{j_n}(T^{(n)},X^{(n)})~.$$  
Hence, \eqref{eq:delta_n} and~\eqref{eq:exch} imply that $P_{n,j_n} \leq \alpha + o(1)$ under the null hypothesis.  To complete the argument, note that $S_n^{\rm obs} = s_{n,I_1}$ and 
\begin{equation} \label{eq:real}
 P\{s_{n,I_1} > \hat c_n(1 - \alpha, \hat \beta_n) |  T^{(n)}, X^{(n)}\} = 
 \sum_{1 \leq i \leq n} \lambda_{i,n} P_{n, i}~,
\end{equation}
where $\lambda_{i,n} = P\{I_1 = i | T^{(n)}, X^{(n)}\}$.  Similar arguments using $\overline i_n = \arg \max_{i \in \mathbb N} P_{n, i}$ in place of $\underline i_n$ show that \eqref{eq:real} is bounded above by $P_{n, j_n} + o(1)$ under the null hypothesis.  It now follows immediately that \eqref{eq:real} is no greater than $\alpha + o(1)$ under the null hypothesis.  

Sufficient conditions for \eqref{eq:delta_n} include settings in which, among other conditions, $\hat c_n(1 - \alpha, \hat \beta_n)$ converges in probability to a constant and $\snn |T^{(n)},X^{(n)}$ is exchangeable under the null hypothesis.  While it is a strong requirement to expect $\snn |T^{(n)}, X^{(n)}$ to be exchangeable under the null hypothesis, it may be approximately so in some instances, especially for a judicious choice of the test statistic.  In such instances, we expect the resulting test to have rejection probability approximately no greater than the nominal level under the null hypothesis. We illustrate this property via a small simulation study in Section \ref{sec:sim_misspecify_DR}.

\subsection{Relationship to Other Methods} \label{sec:firpo}

As mentioned previously, if \eqref{eq:conddistr}, the conditional distribution of the identity of the first adopter, $I_1$, is uniform, then the test defined in \eqref{eq:test} equals the test proposed by \cite{abadie2010synthetic}.  Our testing procedure is also closely related to the analysis of \cite{firpo2018synthetic}, who discuss the properties of this test further and explore its robustness to deviations from the assumption that it is uniform using the sensitivity analysis framework of
 \citet{rosenbaum1987sensitivity,rosenbaum2002observational}.  For a given choice of $S_n$, they first parameterize the weights $\omega_i$ in \eqref{eq:pval} as 
\begin{equation} \label{eq:sens}
\frac{\exp(v_i \phi)}{\sum_{1 \leq k \leq n} \exp(v_k \phi)} ~,
\end{equation}
where $v_i \in \{0,1\}$ is unobserved for all $i \in \mathbb N$ and $\phi \in \Real$ and, for an adversarial choice of $v = (v_i : i \in \mathbb N)$, find the smallest value of $\phi$ that results in a $p$-value that differs in a meaningful way from the same $p$-value when $\phi = 0$.  Despite the apparent similarity between the weights in \eqref{eq:sens} with those in \eqref{eq:pin}, we emphasize that \cite{firpo2018synthetic} do not derive the weights in \eqref{eq:sens} from more primitive assumptions on the time at which each unit adopts treatment like we do here.  In this way, our results provide an alternative viewpoint on their specific parameterization.

\section{Simulations} \label{sec:sims}

In this section, we explore the finite-sample behavior of our proposed testing procedure with a small simulation study.  We first consider in Sections \ref{sec:well_spec} a situation in which the model for the time at which units adopt treatment is correctly specified; we then consider in Section \ref{sec:mis_spec} a situation in which this model is incorrectly specified.  Finally, in Section \ref{sec:sim_misspecify_DR}, we illustrate the robustness property of our testing procedure to misspecification described in Section \ref{sec:robustness}.

\subsection{Correct Specification}\label{sec:well_spec}

For $i \in \mathbb N$ and $t \in \mathbb T$, we assume that $X_{i,t} = X_i$ and $Y_{i,t}(s)$ satisfy \eqref{eq:restrictive} with
\begin{eqnarray*} \label{eq:sim1}
\tilde Y_{i,t}(0) &=& \rho \tilde Y_{i,t-1}(0) + \delta \sqrt t + \gamma X_i +  \epsilon_{i,t} \\
\tilde Y_{i,t}(1) &=& \tau + \tilde Y_{i,t}(0)
\end{eqnarray*}
with $\tilde Y_{i,0}(0) = 0$ for all $i \in \mathbb N$. We further assume that $(T_i,X_i), i \in \mathbb N$ are i.i.d.\ with each $X_i \sim U(-10,10)$ and $T_i|X_i \sim \text{Exp}(\lambda_i)$ with $\lambda_i = \exp(X_i\beta)$ and $\beta = 1$.  It is straightforward to verify that this distribution of $T_i$ satisfies Assumption \ref{ass:prop} with baseline hazard equal to the hazard function of an exponential distribution with parameter equal to one. Finally, independently of $(T^{(n)},X^{(n)})$, $\epsilon_{i,t}, i \in \mathbb N, t \in \mathbb T$ are i.i.d.\ with each $\epsilon_{i,t} \sim N(0,\sigma^2)$.  It therefore follows that Assumption \ref{ass:unconf} is satisfied as well.  Note that under these assumptions the null hypothesis \eqref{eq:null} is satisfied if and only if $\tau = 0$.

We consider three tests.  We first consider our proposed test with $S_n$ in \eqref{eq:tstat} given by a difference-in-differences test statistic:
\begin{eqnarray}
&& \frac{1}{\tmax-\lceil T_{(1)}\rceil } \sum_{T_{(1)} < t \leq \tmax} 
\left(Y_{I_1, t} - \frac{1}{n-1} \sum_{i \in \mathbb N : i\neq I_1} Y_{i, t}\right) \nonumber \\
&& \hspace{4cm} - \frac{1}{\lfloor T_{(1)}\rfloor}\sum_{1 \leq t \le  T_{(1)}} 
\left(Y_{I_1, t} - \frac{1}{n-1} \sum_{i \in \mathbb N : i\neq I_1} Y_{i, t}\right)~.\label{eq:sim1_tstat}
\end{eqnarray}
As mentioned in Section \ref{sec:choice}, another possible choice is the synthetic control test statistic defined in \eqref{eq:synth}, but this simpler choice of test statistic facilitates computation as well as some analytical calculations we present below in Remark \ref{remark:whyworks}.  We emphasize that in our empirical application in Section \ref{sec:empirical} we employ a synthetic control test statistic.  In addition to this test, we also consider the infeasible test which treats $\beta$ as known.  In order to distinguish these two tests, we refer to them as the `feasible' and `infeasible' tests, respectively.  Finally, we also consider the test proposed by \cite{abadie2010synthetic}. In our discussion, we refer to this test as the `uniform' test since it corresponds to our testing procedure with $\omega_i = 1/n$ for all $i \in \mathbb N$ and \eqref{eq:sim1_tstat} as the test statistic. Despite this correspondence, we emphasize that the test of~\cite{abadie2010synthetic} was 
mainly proposed for settings in which a single unit adopts treatment, which differs from our 
setting in which multiple units adopt treatment over time.

In all of our simulations, the nominal level is set to $\alpha = 0.05$ and rejection probabilities are computed using 100,000 replications.  We set $\rho = 0.2$ and $\sigma = 0.2$, but vary $\gamma \in \{0.0, 0.5, 1.0, 2.0, 5.0\}$.  We do not report results for different values of $\delta$ since the specific choice $S_n$ above is invariant with respect to different values of $\delta$.  We vary $n \in \{25, 50, 100\}$ and define $\tmax$ so that $P\{T_i \le \tmax\}=0.15$ to guarantee that some units adopt treatment in the sample. 
 
In Table \ref{tab:sim_valid}, we examine the behavior of the three tests under the null hypothesis, i.e., with $\tau = 0$.  As expected, after accounting for simulation error, the infeasible test has rejection probability no greater than the nominal level for any value of $\gamma$.  In accordance with Theorem \ref{thm:main}, the feasible test has rejection probability that is close to that of the infeasible test and therefore does not exceed the nominal level by a meaningful amount for any value of $\gamma$.  When $\gamma = 0$, i.e., when $X_i$ does not enter \eqref{eq:sim1_tstat}, the uniform test also has rejection probability no greater than the nominal level, but, for all other values of $\gamma$, the test exhibits rejection probabilities that exceed the nominal level by a considerable amount.  For a more detailed discussion of the behavior of these tests when $\gamma = 0$, see Remark \ref{remark:whyworks} below. 

\renewcommand{\arraystretch}{1.15}
\begin{table}[t!]
\centering
\begin{tabular}{cc || ccc}
  \hline
$n$ & $\gamma$ & uniform & feasible & infeasible \\ 
  \hline \hline
25 & 0.00 & 4.96 & 5.11 & 5.00 \\ 
& 0.50 & 14.26 & 4.62 & 4.97 \\ 
 & 1.00 & 17.06 & 4.48 & 5.02 \\ 
 & 2.00 & 18.32 & 4.46 & 5.06 \\ 
& 5.00 & 19.40 & 4.46 & 5.07 \\ 
\hline
  50 & 0.00 & 5.10 & 5.03 & 5.11 \\ 
& 0.50 & 14.30 & 4.82 & 4.98 \\ 
 & 1.00 & 17.07 & 4.77 & 4.95 \\ 
& 2.00 & 19.22 & 4.58 & 4.91 \\ 
 & 5.00 & 20.73 & 4.77 & 5.04 \\ 
\hline
  100 & 0.00 & 4.88 & 5.07 & 5.00 \\ 
 & 0.50 & 14.01 & 4.85 & 4.99 \\ 
& 1.00 & 17.12 & 4.79 & 5.06 \\ 
& 2.00 & 19.71 & 4.85 & 4.95 \\ 
& 5.00 & 21.00 & 4.79 & 5.11 \\ 
   \hline
\end{tabular}
\caption{Rejection rates~(\%) of `uniform', `feasible' and `infeasible' tests of \eqref{eq:null} under the null hypothesis, i.e., when $\tau = 0$.}
\label{tab:sim_valid}
\end{table}

In Table \ref{tab:sim_power}, we examine the behavior of the feasible and infeasible tests under the alternative hypothesis, i.e., $\tau\in\{0.25, 0.5\}$.  In light of the considerable over-rejection under the null hypothesis observed in Table \ref{tab:sim_valid}, we do not examine the uniform test any further.  As expected, after accounting for simulation error, the rejection probabilities for both tests increase as $\tau$ and $n$ increase.  We also see that the rejection probabilities of the feasible and infeasible tests are nearly equal.  Finally, we note that power for both tests decreases as $\gamma$ increases.  This phenomenon can be attributed to the increased variability in $S_n$ for larger values of $\gamma$, which may be seen using calculations like those described in Remark \ref{remark:whyworks} below.

\renewcommand{\arraystretch}{1.15}
\begin{table}[t!]
\centering
\begin{tabular}{cc || cc  | cc}
\hline
&&  \multicolumn{2}{c}{$\tau = 0.25$}  & \multicolumn{2}{c}{$\tau = 0.5$}\\
$n$ & $\gamma$ & feasible  & infeasible & feasible  & infeasible\\ 
  \hline
 25& 0.00 & 19.18 & 19.32 & 24.05 & 24.17\\ 
   & 0.50 & 15.01 & 14.89 & 20.35 & 20.34\\ 
  & 1.00 & 11.48 & 11.91 & 16.38 & 16.77\\ 
   & 2.00 & 8.68 & 9.20 & 12.43 & 12.67\\ 
   & 5.00 & 6.00 & 6.56 & 7.96 & 8.56 \\ 
\hline
 50 & 0.00 & 32.05 & 31.99 & 38.80 & 38.47\\ 
   & 0.50 & 26.98 & 26.93 & 34.90 & 35.28 \\ 
   & 1.00 & 21.96 & 22.07 & 30.73 & 30.91 \\ 
  & 2.00 & 16.71 & 16.59 & 24.50 & 24.55 \\ 
  & 5.00 & 10.10 & 10.43 & 15.81 & 16.01 \\ 
 \hline
 100 & 0.00 & 48.00 & 48.34 & 58.94 & 58.89  \\ 
  & 0.50 & 42.09 & 42.48  & 54.90 & 54.90\\ 
  & 1.00 & 36.11 & 36.25 & 49.49 & 49.58 \\ 
  & 2.00 & 27.63 & 27.70 & 41.48 & 41.51 \\ 
   & 5.00 & 16.80 & 17.12 & 27.88 & 28.28 \\ 
 \hline
\end{tabular}
  \caption{Rejection rates~(\%) of `feasible' and `infeasible' tests of \eqref{eq:null} under the alternative hypothesis, i.e., when $\tau = 0.25$ or $\tau = 0.50$.}
\label{tab:sim_power}
\end{table}
\color{black}

\begin{remark} \label{remark:whyworks}
({\it Behavior of tests when \texorpdfstring{$\gamma = 0$}{}}). A modest amount of calculation shows that under the distributional assumptions described above the test statistic in \eqref{eq:sim1_tstat} may be written, for a suitable function $g$, as
\begin{equation} \label{eq:sim4}
S_n = \frac{n}{n-1} \gamma (\rho_+  - \rho_-) (\bar X_n- X_{I_1}) + \eta_{I_1}
- g(\eta)~,
\end{equation}
under the null hypothesis, where 
\begin{eqnarray}
\bar X_n &=& \frac{1}{n} \sum_{1 \leq i \leq n} X_i~, \nonumber \\
\rho_+ &=& \frac{1}{\tmax- \lceil T_{(1)} \rceil} \sum_{T_{(1)} < t \leq \tmax} \rho^t/(1-\rho)~, \label{eq:rplus}\\
\rho_- &=& \frac{1}{\lfloor T_{(1)} \rfloor} \sum_{1 \leq t < T_{(1)}} \rho^t/(1-\rho) \label{eq:rminus}~,
\end{eqnarray}
and $\eta = (\eta_1, \ldots, \eta_n)$ is an exchangeable normal random variable that is independent of $T_{(1)}$ and $X^{(n)}$.  We note, in particular, that the difference-in-differences statistic eliminates any dependence on $\delta$, but the effect of $\gamma$ persists.  When $\gamma = 0$, the effect of $I_1$ and $X_i$ are both eliminated from \eqref{eq:sim4}.  Using arguments like those in Section \ref{sec:robustness}, it follows that when $\gamma = 0$ any of the tests considered in this section will have rejection probability under the null hypothesis no greater than the nominal level.  Of course, when $\gamma \neq 0$, this need not be the case, as shown by the simulation results presented above.
\end{remark}

\newcommand{\expit}{\ell}
\subsection{Misspecification} \label{sec:mis_spec}

In this section, we explore the robustness of our proposed test to misspecification of Assumption \ref{ass:prop}, i.e., the model for the times at which units adopt treatment.  To this end, we retain our specification in Section \ref{sec:well_spec}, but, importantly, replace the assumption governing the distribution of $T_i|X_i$ with the following specification:
\begin{equation} \label{eq:aft}
T_i = \exp(-m(X_i) + \zeta_i)~,
\end{equation}
where $m(x) = 1 - \ell(2k_1(x + \theta)) + \ell(2k_2(x - \theta))$ with $\ell(u) = \frac{\exp(u)}{1 + \exp(u)}$ (i.e., $\ell(u)$ is the logistic function), $\zeta_i, i \in \mathbb N$ i.i.d.\ with each $\zeta_i \sim N(0,0.4^2)$, and $k_1$, $k_2$ and $\theta$ are non-negative parameters.  We note that this distribution is known in the survival analysis literature as the accelerated failure time model \citep[Section 2.3.3]{kalbfleisch2011statistical}.  It is easily checked that it does not satisfy Assumption \ref{ass:prop} except when $k_1 = k_2 = 0$.

In order to help understand the results of our simulation below, we note some features of the accelerated failure time model above.  If $k_1 = k_2 = 0$, then the $T_i$ are independent of $X_i$ and thus i.i.d.\ conditional on $X^{(n)}$.  It follows that all units are equally likely to be the first adopter.  In fact, as mentioned previously, in this case Assumption \ref{ass:prop} is satisfied with $\beta = 0$.  If either $k_1 > 0$ or $k_2 > 0$, on the other hand, then this is no longer the case.  In particular, when $k_1 > 0$ and $k_2 = 0$, units with low values of $X_i$ are more likely to be the first adopter; when $k_1 = 0$ and $k_2 > 0$, units with high values of $X_i$ are more likely to be the first adopter; and when $k_1 > 0$ and $k_2 > 0$, units with low or high values of $X_i$ are more likely to be the first adopter.

We consider the same three tests defined in Section \ref{sec:well_spec} with the understanding that the `feasible' test in this instance is computed using approximations of $P\{I_1 = i| T_{(1)}, X^{(n)}\}$ computed using simulation and \eqref{eq:aft}.  As before, in all of our simulations, the nominal level is set to $\alpha = 0.05$ and rejection probabilities are computed using 100,000 replications.  We set $\rho = 0.2$, $\sigma = 0.2$ and $\theta = 8$, but vary $k_1 \in\{0, 1, 2\}$, $k_2\in\{0, 1, 2\}$, and $\gamma\in\{0, 2, 5\}$.  In all cases $S_n$ is still given by \eqref{eq:sim1_tstat}, so we do not report results for different values of $\delta$ for the same reason as before.  Finally, we vary $n \in\{25,50\}$ and define $\tmax$ so that $P\{T_i \le \tmax\}=0.15$. 

In Table \ref{tab:sim_misspecify1_full}, we examine the behavior of the three tests under the null hypothesis, i.e., with $\tau = 0$.  We make the following observations.  First, in all cases, the `infeasible' test has rejection probability no greater than the nominal level after accounting for simulation error.  Second, in accordance with the discussion in Section \ref{sec:robustness}, when $\gamma = 0$, all three tests have rejection probability no greater than the nominal level after accounting for simulation error.  Third, when $k_1 = k_2 = 0$, all three tests have rejection probability no greater than the nominal level.  This phenomenon is a consequence of the discussion above that in this instance all units are equally likely to be the first adopter.  Finally, when either $k_1 > 0$ or $k_2 > 0$ and $\gamma \neq 0$, both the `uniform' and `feasible' test exhibit over-rejection in at least some instances, but, in general, we see that the latter exhibits less over-rejection than the former.  Indeed, this pattern is especially evident in situations with $0 = \min\{k_1,k_2\} < \max\{k_1,k_2\}$, i.e., exactly one $k_j = 0$ and exactly one $k_j > 0$.  We believe this phenomenon may be attributable to the `feasible' test's ability to adapt to a limited degree to the fact that not all units are equally likely to be the first adopter.

\renewcommand{\arraystretch}{.96}
\begin{table}[t!]
\centering
\begin{tabular}{ccc|| ccc|ccc}
  \hline
\multicolumn{3}{c}{} & \multicolumn{3}{|c|}{$n = 25$} & \multicolumn{3}{|c}{$n = 50$} \\ \hline
$k_1$ & $k_2$ & $\gamma$ & uniform & feasible & infeasible & uniform & feasible & infeasible \\
\hline \hline
0.00 & 0.00 & 0.00 & 5.10 & 5.08 & 5.02 & 5.01 & 4.98 & 4.99 \\ 
  0.00 & 0.00 & 2.00 & 4.99 & 4.77 & 4.92 & 5.01 & 4.99 & 5.02 \\ 
   0.00 & 0.00 & 5.00 & 4.93 & 4.40 & 5.04 & 4.94 & 4.69 & 4.98 \\ 
\hline
   1.00 & 0.00 & 0.00 & 5.04 & 3.58 & 5.07 & 4.88 & 3.24 & 4.95 \\  
   1.00 & 0.00 & 2.00 & 8.01 & 4.61 & 4.97 & 9.73 & 4.84 & 4.95 \\  
   1.00 & 0.00 & 5.00 & 12.90 & 6.55 & 5.08 & 16.88 & 7.78 & 5.11 \\  
\hline
   2.00 & 0.00 & 0.00 & 5.03 & 3.28 & 5.03 & 5.10 & 3.36 & 5.01 \\  
   2.00 & 0.00 & 2.00 & 8.32 & 4.57 & 4.97 & 10.25 & 4.93 & 4.97 \\  
   2.00 & 0.00 & 5.00 & 13.98 & 6.85 & 5.07 & 17.63 & 7.98 & 4.98 \\  
\hline
   0.00 & 1.00 & 0.00 & 5.03 & 3.59 & 5.04 & 5.07 & 3.44 & 4.93 \\  
   0.00 & 1.00 & 2.00 & 8.11 & 4.57 & 5.08 & 9.87 & 4.77 & 4.88 \\ 
   0.00 & 1.00 & 5.00 & 13.14 & 6.50 & 5.16 & 16.81 & 7.63 & 5.03 \\  
\hline
   1.00 & 1.00 & 0.00 & 5.04 & 4.51 & 5.00 & 4.98 & 4.58 & 4.92 \\  
   1.00 & 1.00 & 2.00 & 8.42 & 7.57 & 5.14 & 9.98 & 9.62 & 5.10 \\  
   1.00 & 1.00 & 5.00 & 14.51 & 13.50 & 5.21 & 17.62 & 17.61 & 5.06 \\  
\hline
   2.00 & 1.00 & 0.00 & 4.98 & 4.50 & 4.96 & 4.97 & 4.55 & 5.01 \\  
   2.00 & 1.00 & 2.00 & 8.70 & 7.76 & 5.13 & 10.06 & 9.68 & 5.06 \\  
   2.00 & 1.00 & 5.00 & 14.41 & 13.70 & 5.14 & 17.90 & 18.17 & 5.17 \\  
\hline
   0.00 & 2.00 & 0.00 & 4.99 & 3.40 & 5.10 & 5.04 & 3.36 & 5.04 \\ 
   0.00 & 2.00 & 2.00 & 8.40 & 4.48 & 5.03 & 10.28 & 4.81 & 4.82 \\  
   0.00 & 2.00 & 5.00 & 13.63 & 6.78 & 4.98 & 17.57 & 7.93 & 5.08 \\ 
\hline
   1.00 & 2.00 & 0.00 & 4.95 & 4.48 & 4.96 & 5.03 & 4.62 & 5.11 \\  
   1.00 & 2.00 & 2.00 & 8.49 & 7.64 & 5.12 & 10.31 & 9.98 & 4.98 \\  
   1.00 & 2.00 & 5.00 & 14.44 & 13.50 & 5.24 & 17.86 & 17.97 & 5.06 \\  
\hline
   2.00 & 2.00 & 0.00 & 4.98 & 4.45 & 5.11 & 4.94 & 4.57 & 4.99 \\  
   2.00 & 2.00 & 2.00 & 8.59 & 7.91 & 5.05 & 10.48 & 10.18 & 5.03 \\ 
   2.00 & 2.00 & 5.00 & 14.62 & 13.89 & 5.12 & 17.81 & 17.79 & 4.99 \\  
\hline
\end{tabular}
\caption{Rejection rates~(\%) of `uniform', `feasible' and `infeasible' tests of \eqref{eq:null} under the null hypothesis, i.e., when $\tau = 0$, and under misspecification.}
\label{tab:sim_misspecify1_full}
\end{table}

\subsection{Robustness}\label{sec:sim_misspecify_DR}


In this section, we illustrate the robustness property of our testing procedure to misspecification as described in Section \ref{sec:robustness}.  To that end, we use the same simulation design as in the preceding section, but we change the test statistic so that the exchangeability property described in Section \ref{sec:robustness} holds approximately.  To motivate our proposed test statistic, consider first
\begin{equation}\label{eq:tstat_DR}
S_n^{\rm R} = S_n - \frac{n}{n-1} \gamma (\rho_+  - \rho_-) (\bar X_n- X_{I_1})~,
\end{equation}
where $S_n$ is the difference-in-differences statistic defined in \eqref{eq:sim1_tstat} and $\rho_+$ and $\rho_-$ are defined in \eqref{eq:rplus}--\eqref{eq:rminus}.  Using the derivations described in Remark \ref{remark:whyworks}, it follows immediately that this choice of test statistic satisfies the desired exchangeability property.  Of course, this test statistic is infeasible because it depends on the unknown parameters $\gamma$, $\rho_+$ and $\rho_-$.  We therefore consider instead
\begin{equation}\label{eq:tstat_DR_f}
S_n^{\mathrm{R, f}} = S_n - \frac{n}{n-1} \hat \gamma_n (\hat{\rho}_{n,+}  - \hat{\rho}_{n,-}) (\bar X_n- X_{I_1})~,
\end{equation}
in which these quantities are replaced by their natural estimators; see Appendix \ref{appendix:sims} for exact definitions of these estimators.  The discussion in Section \ref{sec:robustness} suggests that for such a choice of test statistic our testing procedure may be robust to misspecification of Assumption \ref{ass:prop}.

We consider the same three tests defined in Section \ref{sec:mis_spec} with the choice of test statistic specified in \eqref{eq:tstat_DR_f}.  For comparison, we also include the same three tests with the choice of test statistic specified in \eqref{eq:tstat_DR}.  To facilitate comparison with the results in Table \ref{tab:sim_misspecify1_full}, no other parameters of the simulation are changed.  For brevity, however, we only present results for $n = 50$.  The results of this exercise are displayed in Table \ref{tab:sim_misspecify_DR}.  The key observation is that now all three tests have rejection probability no greater than the nominal level after accounting for simulation error.
  This feature agrees with the discussion in Section \ref{sec:robustness} and contrasts sharply with the results in Table \ref{tab:sim_misspecify1_full}, in which the `uniform' and `feasible' tests both exhibited over-rejection in some cases to varying extents.

\renewcommand{\arraystretch}{1.1}
\begin{table}[t!]
\centering
\begin{tabular}{cccc || ccc | ccc}
  \hline
  & & & & \multicolumn{3}{c}{$S_n^{\mathrm{R, f}}$} & \multicolumn{3}{c}{$S_n^{\mathrm{R}}$} \\ \hline
$n$ & $k_1$ & $k_2$ & $\gamma$ & uniform & feasible & infeasible & uniform & feasible & infeasible \\ 
  \hline \hline
50 & 0.00 & 0.00 & 0.00 & 4.93 & 4.92 & 4.93 & 4.99 & 5.00 & 5.01 \\ 
   & 0.00 & 0.00 & 2.00 & 5.02 & 4.98 & 5.03 & 5.04 & 5.01 & 5.03 \\ 
   & 0.00 & 0.00 & 5.00 & 4.96 & 4.99 & 4.93 & 5.04 & 5.02 & 4.97 \\
  \hline
   & 1.00 & 0.00 & 0.00 & 5.31 & 3.57 & 4.94 & 5.04 & 3.42 & 4.99 \\ 
   & 1.00 & 0.00 & 2.00 & 4.89 & 3.32 & 5.11 & 5.02 & 3.51 & 5.03 \\ 
   & 1.00 & 0.00 & 5.00 & 4.97 & 3.39 & 4.99 & 4.89 & 3.23 & 4.96 \\ 
    \hline
   & 2.00 & 0.00 & 0.00 & 5.14 & 3.25 & 5.12 & 5.02 & 3.32 & 5.08 \\ 
   & 2.00 & 0.00 & 2.00 & 4.99 & 3.30 & 4.99 & 4.90 & 3.27 & 4.96 \\ 
   & 2.00 & 0.00 & 5.00 & 4.98 & 3.30 & 5.05 & 4.97 & 3.18 & 4.95 \\ 
      \hline
   & 0.00 & 1.00 & 0.00 & 5.08 & 3.44 & 5.01 & 5.07 & 3.36 & 4.85 \\ 
   & 0.00 & 1.00 & 2.00 & 5.01 & 3.41 & 5.01 & 5.18 & 3.47 & 5.06 \\ 
   & 0.00 & 1.00 & 5.00 & 5.02 & 3.35 & 5.03 & 4.98 & 3.33 & 5.00 \\ 
        \hline
   & 1.00 & 1.00 & 0.00 & 5.26 & 4.88 & 4.97 & 5.04 & 4.62 & 5.03 \\ 
   & 1.00 & 1.00 & 2.00 & 5.04 & 4.64 & 5.03 & 5.02 & 4.72 & 5.05 \\ 
   & 1.00 & 1.00 & 5.00 & 5.08 & 4.68 & 5.06 & 4.97 & 4.64 & 4.95 \\ 
          \hline
   & 2.00 & 1.00 & 0.00 & 5.23 & 4.71 & 5.09 & 4.98 & 4.53 & 5.02 \\ 
   & 2.00 & 1.00 & 2.00 & 4.98 & 4.53 & 4.96 & 5.02 & 4.68 & 5.20 \\ 
   & 2.00 & 1.00 & 5.00 & 5.13 & 4.67 & 5.18 & 4.96 & 4.62 & 4.97 \\ 
            \hline
   & 0.00 & 2.00 & 0.00 & 5.19 & 3.27 & 5.06 & 4.93 & 3.25 & 4.95 \\ 
   & 0.00 & 2.00 & 2.00 & 5.07 & 3.43 & 5.08 & 5.03 & 3.33 & 5.00 \\ 
   & 0.00 & 2.00 & 5.00 & 4.98 & 3.32 & 5.02 & 4.88 & 3.18 & 4.92 \\ 
              \hline
   & 1.00 & 2.00 & 0.00 & 5.29 & 4.88 & 5.07 & 4.89 & 4.58 & 5.02 \\ 
   & 1.00 & 2.00 & 2.00 & 4.91 & 4.43 & 4.96 & 4.96 & 4.50 & 5.04 \\ 
   & 1.00 & 2.00 & 5.00 & 4.85 & 4.52 & 4.98 & 5.00 & 4.69 & 5.04 \\ 
                \hline
   & 2.00 & 2.00 & 0.00 & 5.11 & 4.71 & 4.89 & 4.99 & 4.52 & 4.98 \\ 
   & 2.00 & 2.00 & 2.00 & 4.98 & 4.66 & 5.00 & 5.12 & 4.68 & 5.04 \\ 
   & 2.00 & 2.00 & 5.00 & 5.14 & 4.68 & 5.07 & 4.90 & 4.52 & 4.98 \\ 
   \hline
  \end{tabular}
  \caption{Rejection rates~(\%) of `uniform', `feasible' and `infeasible' tests of \eqref{eq:null} under the null hypothesis, i.e., when $\tau = 0$, and  with test statistics defined in \eqref{eq:tstat_DR} and \eqref{eq:tstat_DR_f} under misspecification of the treatment adoption model.}
    \label{tab:sim_misspecify_DR}
\end{table}

\color{black}
\clearpage

\section{Empirical Application} \label{sec:empirical}

\begin{figure}[t!]
\centering
\includegraphics[width=.90\textwidth]{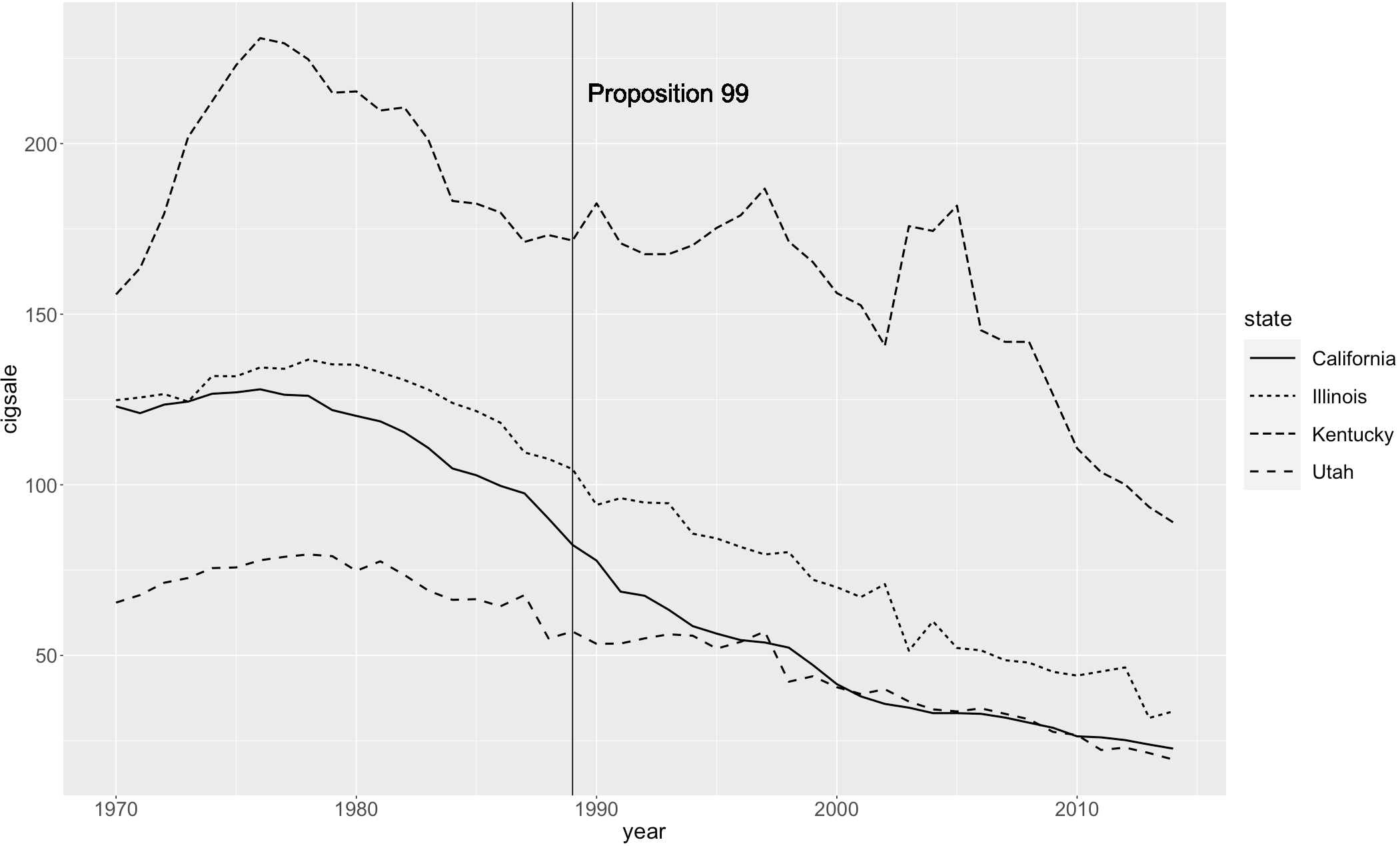}
\caption{Annual cigarette packet sales per capita from 1970 to 2015 in select states. Proposition 99 was adopted by California in 1989.}
\label{fig1}
\end{figure}

In this section, we apply our proposed test to revisit the analysis in \cite{abadie2010synthetic} of the effect of tobacco legislation on smoking prevalence.  We recall that \cite{abadie2010synthetic} was motivated by California's adoption in 1989 of Proposition 99, a large-scale tobacco control program.  A main component of this legislation was a steep increase in cigarette packet tax by 25 cents, representing an increase in taxes of 250\%~\citep[p.318]{abadie2010synthetic}.  While smoking prevalence declined after this legislation was adopted, it is important to emphasize that this decline happened in the backdrop of nationwide declining smoking prevalence dating at least as far back as the late 1970s.  See Figure~\ref{fig1} for a graphical depiction of these trends for several different states, including California.  In this context, \citet{abadie2010synthetic} used the synthetic control methodology to test the null hypothesis that {\it none} of the decline in smoking prevalence observed in California after 1989 can be attributed to the effects of the Proposition 99. Their analysis rejects this null hypothesis with a $p$-value of $0.026$.

We now describe the application of our feasible test in Algorithm~\ref{algo1} to this setting.  To facilitate comparison with the results in \cite{abadie2010synthetic}, we restrict attention to the same $n =39$ states in their analysis.  These states are indexed by $i \in \mathbb N$.  We index time by $t \in \mathbb T = \{``01/1971",  ``02/1971", \ldots, ``12/2014"\}$, where we have adopted the ``month/year'' format and identify $1$ with $``01/1971"$ and $\tmax$ with $``12/2014"$.  Denote by $Y_{i,t}$ the number of cigarette packets sold in state $i \in \mathbb N$ at time $t \in \mathbb T$.  Finally, let $T_i$ denote the time at which state $i \in \mathbb N$ adopts tobacco legislation.  \citet{orzechowski2014tax} provide a comprehensive record of tobacco tax increases across states during this time period.  In order to resolve any ambiguities, we define this to be the first time taxes on cigarette packets are increased by at least 50\%.  Every state except for Missouri adopt such tobacco legislation in our sample period.  In Appendix \ref{appendix:tobacco_spec}, we examine the robustness of our results to different ways of defining $T_i$.  In addition, while the data permit measurement of $T_i$ up to the day of adoption, we simply record the month of adoption.  In particular, $T_{(1)} = ``01/1989"$.  We emphasize, however, that no two states adopted tobacco legislation during the same month, so this is immaterial.

We employ the test statistic defined in \eqref{eq:synth}.  
The weights in the construction of $\hin$ are computed as follows
\begin{equation}\label{eq:synth_w}
w^\ast = \arg\min_{w} \sum_{t\in\mathbb T : t < T_{(1)}} \left (Y_{I_1,t} - \sum_{j \in \mathbb N : j \neq I_1} w_j Y_{j, t} \right )^2~,
\end{equation}
where the possible values of $w$ are understood to be in the $n$-dimensional simplex with $w_{I_1}$ restricted to be equal to zero.
In order to complete the description of our testing procedure, we assume that the times at which states adopt tobacco legislation is governed by a Cox proportional hazards model as described in Assumption \ref{ass:prop}, which requires, in particular, a specification of the covariates.  We include the following covariates: per capita income (in logs), average price levels, fraction of population that are youth (with youth defined to be people with ages between 15 and 24), the unemployment level, and the fraction of state legislators that are Democrats.  We note that these variables are not time-invariant and change on a yearly basis.  
In order to account for the systematic increase in some of these variables over time, we first de-trend these variables using a common linear trend across states.  We emphasize that our test remains valid because the main requirement underlying its validity is simply the consistency of $\hat \beta_n$.  Before proceeding, we note that in Appendix \ref{appendix:tobacco_spec} we examine the robustness of our findings to more parsimonious specifications of the covariates.

For our baseline specification defined by the choices above, we compute a $p$-value of 0.044.  We therefore reject the null hypothesis \eqref{eq:null} at conventional significance levels, such as $\alpha = 0.05$.  In order to gain some further insight into this result, it is worthwhile to examine the estimated conditional distribution of the identity of the state that first adopted tobacco legislation, i.e., the distribution of $I_1 | T_{(1)}, X^{(n)}$.  This is presented in Table \ref{tab:synth_ps} for our baseline  specification.   We see that the model of treatment adoption implies that there are nine states that had higher probability than California of being the first to adopt treatment (conditional on the first adoption occurring in January 1989), namely, Nevada, Connecticut, Rhode Island, North Dakota, Maine, Illinois, Wisconsin, Texas, and Nebraska.  In contrast, states such as Kentucky, Missouri or North Carolina, had much lower probability of being the first to adopt treatment (conditional on the first adoption occurring in January 1989).  Indeed, California was more than 34 times as likely to be the first adopter than Kentucky (conditional on the first adoption occurring in January 1989).  These features reflect differences in both the characteristics of these states as well as disparities in the time at which different states adopted treatment.  For instance, Mississippi only introduced such tobacco legislation as late as 2009.

\renewcommand{\arraystretch}{1.15}
\begin{table}[t!]
\centering
\begin{tabular}{l c |lc | lc}
  \hline
State & Prob.\ & State  & Prob.\ & State  & Prob.\ \\ 
 \hline \hline
Nevada & 0.1037 & Arkansas & 0.0250 & Virginia & 0.0086 \\ 
 Connecticut & 0.1018 & Pennsylvania & 0.0225 & Alabama & 0.0081 \\ 
Rhode Island & 0.0681 & Louisiana & 0.0201 & West Virginia & 0.0066 \\ 
North Dakota & 0.0617 & Ohio & 0.0200 & Oklahoma & 0.0037 \\ 
Maine & 0.0605 & Delaware & 0.0186 & South Carolina & 0.0033 \\ 
 Illinois & 0.0580 & Minnesota & 0.0171 & South Dakota & 0.0027 \\ 
 Wisconsin & 0.0517 & Tennessee & 0.0163 & Vermont & 0.0026 \\ 
 Texas & 0.0491 & Montana & 0.0151 & Utah & 0.0023 \\ 
Nebraska & 0.0460 & Idaho & 0.0139 & Iowa & 0.0018 \\ 
{\bf California} & {\bf 0.0440} & Indiana & 0.0134 & North Carolina & 0.0016 \\ 
 New Hampshire & 0.0360 & Kansas & 0.0124 & Missouri & 0.0014 \\ 
 Wyoming & 0.0291 & Georgia & 0.0124 & Kentucky & 0.0013 \\ 
 New Mexico & 0.0279 & Colorado & 0.0111 & Mississippi & 0.0006 \\ 
   \hline
\end{tabular}
\caption{Estimated conditional distribution of the identity of the state that first adopted tobacco legislation, i.e., the distribution of $I_1 | T_{(1)}, X^{(n)}$.}
\label{tab:synth_ps}
\end{table}

We conclude the discussion of our empirical results by noting that had we implemented the test with $\omega_i = 1/n$ we would have computed, like \cite{abadie2010synthetic}, a $p$-value of $0.026$.  This phenomenon simply reflects the fact that for our specification above $S_n(i, Y^{(n)}, T_{(1)}, X^{(n)})$ is largest for $i$ corresponding to California.  As mentioned previously, in Appendix \ref{appendix:tobacco_spec}, we further examine the robustness of this finding to different ways of defining $T_i$ as well as more parsimonious choices of covariates in the Cox proportional hazards model.  We repeat the analysis for more than 30,000 different resulting specifications and find that in the vast majority of specifications, we still reject the null hypothesis at the $\alpha = 0.05$ significance level and in all specifications we reject at the $\alpha = 0.10$ significance level.  In this sense, we believe our analysis largely confirms the findings in \cite{abadie2010synthetic}.  

\section{Concluding Remarks} \label{sec:conclusion}

In this paper, we have developed a method for inference in observational studies with staggered adoption of treatment.  We have focused on testing a null hypothesis that specifies there is no treatment effect for all units and all time periods in a distributional sense. Our proposed test is a randomization test that relies upon an assumption that the time at which each unit adopts treatment follows a Cox proportional hazards model and the availability of multiple units that adopt treatment at different times in order to estimate the parameters of this model consistently.  While the Cox proportional hazards model is semiparametric in nature and widely used for different purposes, its use here is novel and provides a link between the survival analysis literature and these types of settings.  Further exploration of this connection in future research may be fruitful.  The Cox proportional hazards model is nevertheless restrictive.  We have therefore additionally explored robustness properties of our testing procedure to violations of this assumption as well as nonparametric generalizations of our testing procedure.  Finally, we have revisited the analysis in \cite{abadie2010synthetic} of the effect of tobacco legislation on smoking prevalence and added to the evidence presented there of the effect of Proposition 99 on smoking prevalence.

\bibliographystyle{ims}
\bibliography{refs.bib}

\newpage
\appendix
\small
\section{Appendix}

\subsection{Proof of Lemma \ref{lemma:indep}}

Assumption \ref{ass:unconf} and the null hypothesis \eqref{eq:null} imply that $$Y^{(n)} \indep T^{(n)} | X^{(n)}~.$$The desired conclusion \eqref{eq:indep} now follows immediately upon noting that $I_1$ and $T_{(1)}$ are functions of $T^{(n)}$.

\subsection{Proof of Lemma \ref{lemma:prop}}

Note that 
\begin{eqnarray*}
P\{I_1 = i | T_{(1)} = t, X^{(n)} \} &=& P\{T_i = t | T_{(1)} = t, X^{(n)}\} \\
&=& \lim_{\delta \downarrow 0} P\{t \leq T_i \leq t + \delta | t \leq T_{(1)} \leq t + \delta, X^{(n)}\} \\
&=& \lim_{\delta \downarrow 0} \frac{P\{t \leq T_i \leq t + \delta | X_i\} \prod_{1 \leq j \leq n : j \neq i} P\{T_j \geq t | X_j\}}{\sum_{1 \leq k \leq n}P\{t \leq T_k \leq t + \delta| X_k\} \prod_{1 \leq j \leq n : j \neq k} P\{T_j \geq t | X_j\} } \\
&=& \lim_{\delta \downarrow 0} \frac{P\{t \leq T_i \leq t + \delta | T_i \geq t, X_i\}}{\sum_{1 \leq k \leq n} P\{t \leq T_k \leq t + \delta | T_k \geq t, X_k\}} \\
&=& \lim_{\delta \downarrow 0} \frac{\frac{1}{\delta}P\{t \leq T_i \leq t + \delta | T_i \geq t, X_i\}}{\sum_{1 \leq k \leq n} \frac{1}{\delta} P\{t \leq T_k \leq t + \delta | T_k \geq t, X_k\}} \\
&=& \frac{\lambda(t)\exp(X_{i,t}'\beta)}{\sum_{1 \leq k \leq n} \lambda(t)\exp(X_{k,t}'\beta)}~,
\end{eqnarray*}
where the first equality follows by inspection, the second equality is understood to be by definition, the third through fifth equalities follow from Bayes' rule, and the sixth equality follows from Assumption \ref{ass:prop}.  The desired conclusion now follows immediately.

\subsection{Proof of Theorem \ref{thm:main}}

Before proceeding, we note that the proof makes use of Lemma A.1 in \cite{romano2012uniform}.  For completeness, we provide a statement of that result here.  Recall that for any c.d.f.\ $F$ on $\mathbf R$ and $\alpha \in [0,1]$, we define $F^{-1}(\alpha) = \inf\{x \in \mathbf R : F(x) \geq \alpha\}$, with the understanding that $F^{-1}(0)$ and $F^{-1}(1)$ are $-\infty$ and $+\infty$, respectively.
\begin{lemma}  \label{lemma:quant}
If $F$ and $G$ are (nonrandom) distribution functions on $\Real$, then we have that:
\begin{enumerate}
\item[(i)] If $\sup_{x \in \mathbf{R}} \{G(x) - F(x)\} \leq \epsilon$, then $G^{-1}(1 - \alpha_2) \geq F^{-1}(1 - (\alpha_2 + \epsilon))~.$
\item[(ii)] If $\sup_{x \in \mathbf{R}} \{F(x) - G(x)\} \leq \epsilon$, then $G^{-1}(\alpha_1) \leq F^{-1}(\alpha_1 + \epsilon)~.$
\end{enumerate}
\noindent Furthermore, if $X \sim F$, it follows that:
\begin{enumerate}
\item[(iii)] If $\sup_{x \in \mathbf{R}} \{G(x) - F(x)\} \leq \epsilon$, then $P\{X \leq G^{-1}(1 - \alpha_2)\} \geq 1 - (\alpha_2 + \epsilon)~.$
\item[(iv)] If $\sup_{x \in \mathbf{R}} \{F(x) - G(x)\} \leq \epsilon$, then $P\{X \geq G^{-1}(\alpha_1)\} \geq 1 - (\alpha_1 + \epsilon)~.$
\item[(v)] If $\sup_{x \in \mathbf{R}} |G(x) - F(x)| \leq \frac{\epsilon}{2}$, then $P\{G^{-1}(\alpha_1) \leq X \leq G^{-1}( 1 - \alpha_2)\} \geq 1 - (\alpha_1 + \alpha_2 + \epsilon)~.$
\end{enumerate}
\noindent If $\hat G$ is a random distribution function on $\mathbf{R}$, then we have further that:
\begin{enumerate}
\item[(vi)] If $P\{\sup_{x \in \mathbf{R}} \{\hat G(x) - F(x)\} \leq \epsilon\} \geq 1 - \delta$, then $P\{ X \leq \hat G^{-1}( 1 - \alpha_2)\} \geq 1 - (\alpha_2 + \epsilon + \delta)~.$
\item[(vii)] If $P\{\sup_{x \in \mathbf{R}} \{F(x) - \hat G(x)\} \leq \epsilon\} \geq 1 - \delta$, then $P\{ X \geq \hat G^{-1}(\alpha_1) \} \geq 1 - (\alpha_1 + \epsilon + \delta)~.$
\item[(viii)] If $P\{\sup_{x \in \mathbf{R}} |\hat G(x) - F(x)| \leq \frac{\epsilon}{2}\} \geq 1 - \delta$, then $P\{\hat G^{-1}(\alpha_1) \leq X \leq \hat G^{-1}( 1 - \alpha_2)\} \geq 1 - (\alpha_1 + \alpha_2 + \epsilon + \delta)~.$
\end{enumerate}
\end{lemma}

\noindent In what follows, we let $\hat \Delta_n = \hat \beta_n - \beta$, and also use $a \lesssim b$ to denote that $a \leq c b$ for some constant $c$.  Before proceeding, note that Assumption \ref{ass:prop}(d) implies that
\begin{equation} \label{eq:boundedbelowabove}
0 < \inf_{x \in \text{supp}(X_{i,t})}\exp(x'\beta) \leq \sup_{x \in \text{supp}(X_{i,t})}\exp(x'\beta) < \infty ~.
\end{equation}
Similarly, Assumptions \ref{ass:prop}(d) and \ref{ass:cons} imply that
\begin{equation} \label{eq:toone}
\sup_{x \in \text{supp}(X_{i,t})}\left |\exp(x'\hat \Delta_n)- 1 \right | = o_P(1)~.
\end{equation}
Now, note that 
\begin{eqnarray*}
&& \sup_{s \in \mathbf R} \left | P\left \{S_n \leq s | Y^{(n)}, T_{(1)}, X^{(n)} \right \} - \sum_{1 \leq i \leq n} \omega_{i,n}(\hat \beta_n) \Ind\left \{S_n\left (i, Y^{(n)}, T_{(1)}, X^{(n)} \right ) \leq s \right \}\right | ~, \\
&\leq& \sum_{1 \leq i \leq n} \left | \omega_{i,n}(\beta) - \omega_{i,n}(\hat \beta_n) \right | \\
&\leq& \sup_{1 \leq i \leq n} \left | \frac{\exp(X_{i,T_{(1)}}'\beta)}{\frac{1}{n}\sum_{1 \leq k \leq n} \exp(X_{k,T_{(1)}}'\beta)} - \frac{\exp(X_{i,T_{(1)}}'\hat \beta_n)}{\frac{1}{n}\sum_{1 \leq k \leq n} \exp(X_{k,T_{(1)}}'\hat \beta_n)} \right | \\
&=& \sup_{1 \leq i \leq n} \left | \frac{\exp(X_{i,T_{(1)}}'\beta) \left (\frac{1}{n}\sum_{1 \leq k \leq n} \exp(X_{k,T_{(1)}}'\hat \beta_n) \right ) - \exp(X_{i,T_{(1)}}'\hat \beta_n) \left ( \frac{1}{n}\sum_{1 \leq k \leq n} \exp(X_{k,T_{(1)}}'\beta) \right )}{\left (\frac{1}{n}\sum_{1 \leq k \leq n} \exp(X_{k,T_{(1)}}'\beta) \right ) \left (\frac{1}{n}\sum_{1 \leq k \leq n} \exp(X_k'\hat \beta_n) \right )} \right | \\
&=& \sup_{1 \leq i \leq n} \left | \frac{\exp(X_{i,T_{(1)}}'\beta) \left ( \left (\frac{1}{n}\sum_{1 \leq k \leq n} \exp(X_{k,T_{(1)}}'\hat \Delta_n)\exp(X_{k,T_{(1)}}'\beta) \right ) - \exp(X_{i,T_{(1)}}'\hat \Delta_n) \left ( \frac{1}{n}\sum_{1 \leq k \leq n} \exp(X_{k,T_{(1)}}'\beta) \right ) \right )}{\left (\frac{1}{n}\sum_{1 \leq k \leq n} \exp(X_{k,T_{(1)}}'\beta) \right ) \left (\frac{1}{n}\sum_{1 \leq k \leq n} \exp(X_{k,T_{(1)}}'\hat \Delta_n) \exp(X_{k,T_{(1)}}'\beta) \right )} \right | \\
&\lesssim& \sup_{1 \leq i \leq n} \left | \frac{ \left ( \left (\frac{1}{n}\sum_{1 \leq k \leq n} \exp(X_{k,T_{(1)}}'\hat \Delta_n)\exp(X_{k,T_{(1)}}'\beta) \right ) - \exp(X_{i,T_{(1)}}'\hat \Delta_n) \left ( \frac{1}{n}\sum_{1 \leq k \leq n} \exp(X_{k,T_{(1)}}'\beta) \right ) \right )}{\frac{1}{n}\sum_{1 \leq k \leq n} \exp(X_{k,T_{(1)}}'\hat \Delta_n) } \right | \\
&=& \sup_{1 \leq i \leq n} \left | \frac{ \left ( \left (\frac{1}{n}\sum_{1 \leq k \leq n} \left (\exp(X_{k,T_{(1)}}'\hat \Delta_n) - 1\right )\exp(X_{k,T_{(1)}}'\beta) \right ) - \left (\exp(X_{i,T_{(1)}}'\hat \Delta_n) - 1 \right) \left ( \frac{1}{n}\sum_{1 \leq k \leq n} \exp(X_{k,T_{(1)}}'\beta) \right ) \right )}{\frac{1}{n}\sum_{1 \leq k \leq n} \left ( \exp(X_{k,T_{(1)}}'\hat \Delta_n) - 1 \right )  + 1} \right | \\
&=& o_P(1)~.
\end{eqnarray*}
where the first inequality exploits \eqref{eq:cdf}, the second inequality exploits \eqref{eq:pin}, the first and second equalities follow by inspection, the third inequality exploits \eqref{eq:boundedbelowabove}, the third equality follows by inspection, and the final equality exploits \eqref{eq:boundedbelowabove}--\eqref{eq:toone}.  The desired result now follows by applying Lemma \ref{lemma:quant}.

\section{Description of Estimators for Section \ref{sec:sim_misspecify_DR}}\label{appendix:sims}

Here, we describe the plug-in estimates of $\gamma,  \rho$ that are used in the 
feasible test statistic, $S_n^{\mathrm{DR, f}}$ of Section~\ref{sec:sim_misspecify_DR}.
The main goal in these estimators is simplicity (and computational efficiency) rather than statistical 
efficiency. It is straightforward to show that all the estimators defined below are consistent 
for an increasing number of units, $n$.

To estimate $\gamma$, note that our outcome model implies that at $t=1$ we have
$$
Y_{i, 1} - Y_{1, 1} = \gamma (X_i -X_1)  + \epsilon_{i, 1} - \epsilon_{1, 1}.
$$
The variance-covariance matrix of errors $\epsilon_{i, 1} - \epsilon_{1, 1}$ is  
equal to $C = \mathbb{I}_n + \mathbb{U}_n$, where $\mathbb{I}$ is the identity matrix 
and $\mathbb{U}$ is the matrix of ones.
Let $y = (Y_{i, 1} - Y_{1, 1} : i \in \mathbb{N})$ and 
$x = (X_i - X_1 : i \in \mathbb{N})$ as column vectors. 
Then, a simple estimator for $\gamma$ is the slope coefficient 
of regressing $C^{-1} y$ on $C^{-1} x$. This estimator, namely $\hat\gamma$, is consistent as $n$ grows.

To estimate $\rho$, let $\mathbf{Y}$ be the $n\times \tmax$ matrix of outcomes and $\mathbf{Y}_1$ be the $(n-1) \times (\tmax -1)$ matrix  
obtained from $\mathbf{Y}$ by removing the row corresponding to the first adopter, $I_1$, 
and also the first column corresponding to $t=1$.
Furthermore, let $\mathbf{Y}_1^-$ be the $(n-1) \times (\tmax -1)$  matrix  
obtained from $\mathbf{Y}$ by removing the row corresponding to the first adopter, $I_1$, 
and  the last column corresponding to $t=\tmax$.
Finally, let $\mathbf{X}$ be the $n \times (\tmax -1)$ matrix where the 
element in $(i, t)$ position is equal to $X_i$, and 
let $\mathbf{X}_1$ be matrix $\mathbf{X}$ with the $I_1$th row removed.
With the same reasoning as above, the particular outcome model in this setting implies that 
a consistent estimator for $\rho$ is obtained 
by regressing 
$D^{-1} (\mathbf{Y}_1 - \hat\gamma \mathbf{X}_1)$ on $D^{-1} \mathbf{Y}_1^-$, 
where $D = \mathbb{I}_{n-1} + \mathbb{U}_{n-1}$ and $\hat\gamma$ is the 
consistent estimator of $\gamma$ described above.
We can use this estimate to obtain estimates $\hat\rho_+$ and $\hat\rho_-$ 
by using the definitions in Remark~\ref{remark:whyworks}.

\section{Additional Specifications for Empirical Application}
\label{appendix:tobacco_spec}

In this section, we examine the robustness of our empirical findings in Section \ref{sec:empirical}.  Specifically, we re-compute our $p$-value for different ways of defining $T_i$ as well as exclusion of some of the six covariates we include in our Cox proportional hazards model.

As mentioned in Section \ref{sec:empirical}, we define $T_i$ to be the first time at which a state increased taxes on cigarette packets by at least 50\%.  While this eliminates any ambiguity, we identify nine states for which an alternative choice of $T_i$ seems reasonable based subjectively on the magnitude or timing of the increase.  These nine states are indicated in bold face in Table \ref{tab:data_spec}.  For each of those states, we indicate in the column labeled `Specification B' the choice of $T_i$ corresponding to our specification in Section \ref{sec:empirical} and the alternative choice in the column labeled `Specification A'.  For all other states, we simply repeat in these two columns the single choice of $T_i$ that we consider.  By considering all possible choices of $T_i$ for these nine states, we obtain $2^9 = 512$ possible specifications of $T_i$.  For each of these specifications, we additionally consider each of the $2^6 = 64$ possible subsets of the six covariates to include in the Cox proportional hazards model.  We therefore obtain in total $512 \times 64 = 32,768$ possible specifications.  In order to facilitate our discussion below, we compute, in addition to the $p$-value for our test, the Akaike Information Criteria (AIC) for each Cox proportional hazards model.  

\renewcommand{\arraystretch}{1.0}
\begin{table}[t!]
\centering
\begin{tabular}{rlcc}
  \hline
 & State & Specification A & Specification B \\ 
  \hline
1 & Alabama & 05/2004 & 05/2004 \\ 
  2 & {\bf Arkansas} & 06/2003 & 02/1993 \\ 
  3 & California & 01/1989 & 01/1989 \\ 
  4 & Colorado & 01/2005 & 01/2005 \\ 
  5 & Connecticut & 04/1989 & 04/1989 \\ 
  6 & {\bf Delaware} & 01/1991 & 08/2003 \\ 
  7 & Georgia & 07/2003 & 07/2003 \\ 
  8 & {\bf Idaho} & 07/1994 & 06/2003 \\ 
  9 & Illinois & 07/1989 & 07/1989 \\ 
  10 & Indiana & 07/2002 & 07/2002 \\ 
  11 & Iowa & 04/2007 & 04/2007 \\ 
  12 & Kansas & 07/2002 & 07/2002 \\ 
  13 & Kentucky & 06/2005 & 06/2005 \\ 
  14 & Louisiana & 08/2002 & 08/2002 \\ 
  15 & {\bf Maine} & 11/1997 & 07/1991 \\ 
  16 & {\bf Minnesota} & 06/1991 & 08/2005 \\ 
  17 & Mississippi & 05/2009 & 05/2009 \\ 
  18 & Missouri & 12/2014 & 12/2014 \\ 
  19 & Montana & 05/2003 & 05/2003 \\ 
  20 & Nebraska & 10/2002 & 10/2002 \\ 
  21 & Nevada & 07/1989 & 07/1989 \\ 
  22 & New Hampshire & 02/1990 & 02/1990 \\ 
  23 & New Mexico & 07/2003 & 07/2003 \\ 
  24 & North Carolina & 09/2005 & 09/2005 \\ 
  25 & North Dakota & 05/1989 & 05/1989 \\ 
  26 & Ohio & 07/2002 & 07/2002 \\ 
  27 & Oklahoma & 01/2005 & 01/2005 \\ 
  28 & Pennsylvania & 08/1991 & 08/1991 \\ 
  29 & {\bf Rhode Island} & 07/1997 & 07/1993 \\ 
  30 & South Carolina & 07/2010 & 07/2010 \\ 
  31 & {\bf South Dakota} & 03/2003 & 07/1995 \\ 
  32 & Tennessee & 07/2002 & 07/2002 \\ 
  33 & Texas & 07/1990 & 07/1990 \\ 
  34 & {\bf Utah} & 07/1991 & 07/1997 \\ 
  35 & Vermont & 07/1995 & 07/1995 \\ 
  36 & Virginia & 09/2004 & 09/2004 \\ 
  37 & West Virginia & 05/2003 & 05/2003 \\ 
  38 & Wisconsin & 05/1992 & 05/1992 \\ 
  39 & {\bf Wyoming} & 07/2003 & 07/1989 \\ 
   \hline
\end{tabular}
\caption{Different specifications for $T_i$.}
\label{tab:data_spec}
\end{table}

\begin{figure}[t!]
\centering
\includegraphics[scale=0.25]{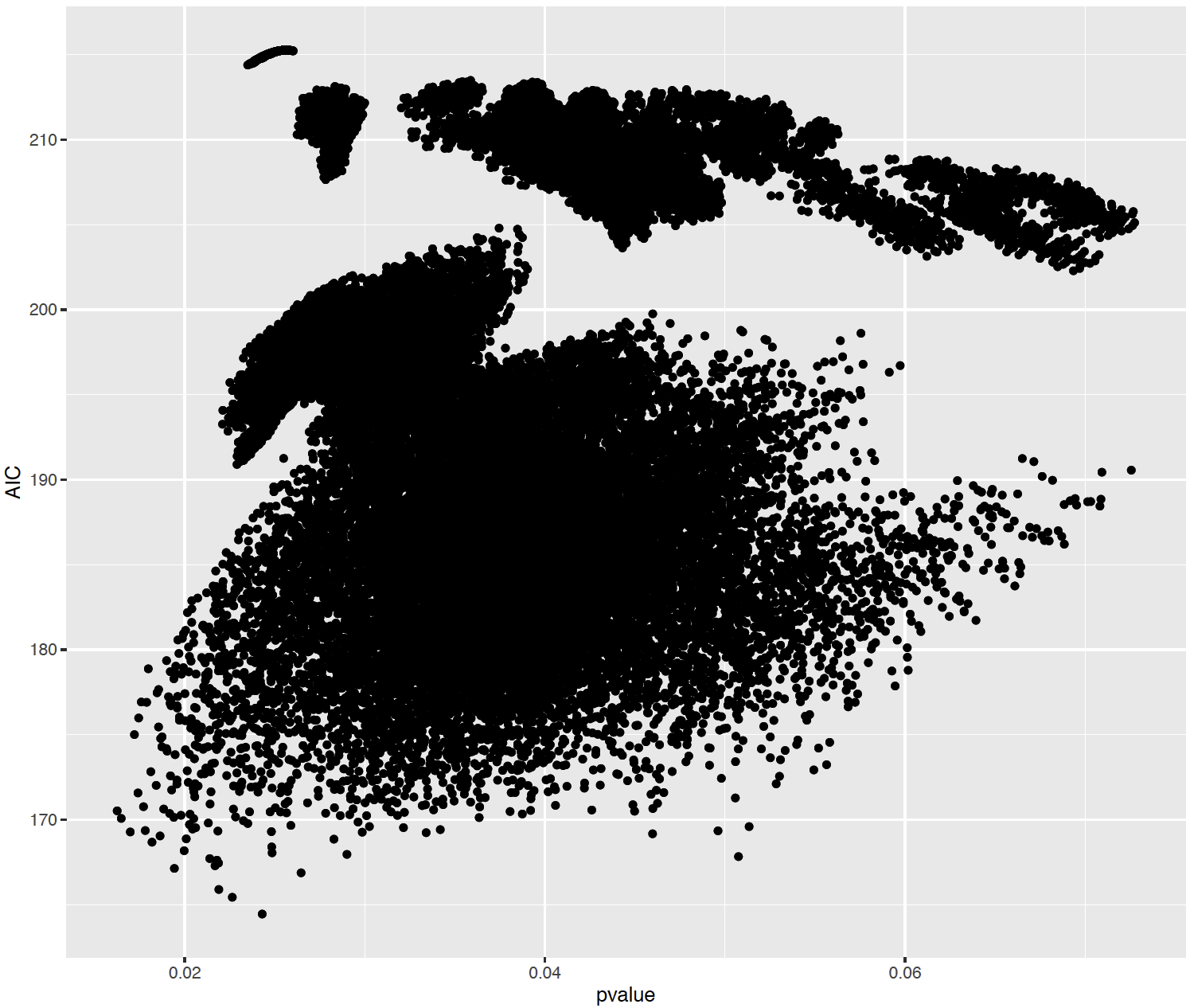}
\caption{Scatter plot of $(p\text{-value}, \text{AIC})$ across all specifications.}
\label{fig:aic}
\end{figure}

In Figure \ref{fig:aic}, we plot for each of these specifications the $(p\text{-value}, \text{AIC})$.  We see that in the vast majority of cases, we continue to reject the null hypothesis at the $\alpha = 0.05$ significance level.  This conclusion is further strengthened if we restrict attention to specifications with better (i.e., lower) values of AIC.  Finally, in all specifications, we reject the null hypothesis at the $\alpha = 0.10$ significance level.  Indeed, the maximum $p$-value across all specifications is 0.072.  In this sense, we find that our findings in Section \ref{sec:empirical} and, by extension, those of \cite{abadie2010synthetic} are remarkably robust.

\end{document}